\documentclass[%
 aip,
 amsmath,amssymb,
 reprint,%
]{revtex4-1}

\usepackage{graphicx}% Include figure files
\usepackage{dcolumn}% Align table columns on decimal point
\usepackage{bm}% bold math
\usepackage[utf8]{inputenc}
\usepackage[T1]{fontenc}
\usepackage{mathptmx}
\usepackage{etoolbox}

\usepackage{color}
\definecolor{red}{rgb}{0.8, 0.0, 0.0}

\makeatletter
\def\@email#1#2{%
 \endgroup
 \patchcmd{\titleblock@produce}
  {\frontmatter@RRAPformat}
  {\frontmatter@RRAPformat{\produce@RRAP{*#1\href{mailto:#2}{#2}}}\frontmatter@RRAPformat}
  {}{}
}%
\makeatother
\begin{document}

\preprint{AIP/123-QED}

\title[Bayesian exploration of the composition space of CuZrAl metallic glasses for mechanical properties]{Bayesian exploration of the composition space of CuZrAl metallic glasses for mechanical properties}
% Force line breaks with \\
\author{Tero Mäkinen}%
\email{tero.j.makinen@aalto.fi}
\affiliation{Aalto University, Department of Applied Physics, P.~O. Box 15600, 00076 Aalto, Espoo, Finland}
\author{Anshul D. S. Parmar}
\affiliation{NOMATEN Centre of Excellence, National Center for Nuclear Research, ul. A. Soltana 7, 05-400 Swierk/Otwock, Poland}
\author{Silvia Bonfanti}
\affiliation{NOMATEN Centre of Excellence, National Center for Nuclear Research, ul. A. Soltana 7, 05-400 Swierk/Otwock, Poland}
\affiliation{Center for Complexity and Biosystems, Department of Physics ‘Aldo Pontremoli’, University of Milan, Milano, Italy}
\author{Mikko J. Alava}
\affiliation{Aalto University, Department of Applied Physics, P.~O. Box 15600, 00076 Aalto, Espoo, Finland}
\affiliation{NOMATEN Centre of Excellence, National Center for Nuclear Research, ul. A. Soltana 7, 05-400 Swierk/Otwock, Poland}

\date{\today}% It is always \today, today,

\begin{abstract}
Designing metallic glasses \textit{in silico} is a major challenge in materials science given their disordered atomic structure and the vast compositional space to explore. Here, we tackle this challenge by finding optimal compositions for target mechanical properties. We apply Bayesian exploration for the CuZrAl composition, a paradigmatic metallic glass known for its good glass forming ability.
We exploit an automated loop with an online database, a Bayesian optimization algorithm, and molecular dynamics simulations. From the ubiquitous 50/50 CuZr starting point, we map the composition landscape changing the ratio of elements and adding aluminium to characterize
the yield stress and the shear modulus. This approach demonstrates with relatively modest effort that the system has an optimal composition window for the yield stress around aluminium concentration $c_{\rm Al} = 15$~\% and zirconium concentration $c_{\rm Zr} = 30$~\%. 
We also explore several cooling rates ("process parameters") and find that the best mechanical properties for a composition result from being most affected by the cooling procedure. 
Our Bayesian approach paves the novel way for the design of metallic glasses with "small data", with an eye toward both future \textit{in silico} design and experimental applications exploiting this toolbox. 
\end{abstract}

\maketitle

\section{Introduction}

Metallic glasses~\cite{greer1995metallic,kruzic2016bulk,suryanarayana2017bulk,greer2023metallic,sohrabi2024manufacturing} (MGs) are attracting ever-increasing interest for their remarkable mechanical properties, such as high strength, hardness, and wear resistance, which make them promising candidates for advanced applications~\cite{eckert2007processing,trexler2010mechanical,egami2013mechanical}. 
These unique properties of MGs stem from their amorphous (disordered) atomic structure, in contrast to the
crystalline (ordered) structure of traditional metals and alloys. 
The amorphous state is achieved by rapidly cooling the
metallic liquid to prevent the atoms from organizing into a crystalline arrangement. This cooling process directly affects a critical feature called glass forming ability~\cite{miracle2010assessment} (GFA), which is the ability of a material to avoid crystallization and maintain a disordered structure during cooling. 
A good GFA is fundamental to making any metallic glass and possibly acquiring unique mechanical properties. 
However, questions remain about how cooling rates~\cite{schroers2013bulk, huang2014effect, yue2018effect, schawe2019existence}, i.e. the variation of temperature with respect to time, and chemical compositions may promote a glassy disordered state (good GFA). In generic terms, while rapid cooling supports GFA, quenching or cooling from the liquid state gives better and better properties the slower the procedure is, though with an increasing risk of crystallization due to the general metastability of the glassy state. 
This account is for ideal MG, however, in the laboratory reality, the glass-making route is a daunting task~\cite{halim2021metallic, zhang20213d}. 

The exploration of metallic glass property landscapes has first and foremost concentrated on the GFA since that is the starting point for considering any particular composition and a preparation route. 
Here and for other properties one faces considerable challenges in coming up with high-throughput experimental approaches for mapping or design/optimization often using machine learning~\cite{doi:10.1126/sciadv.aaq1566,10.1063/5.0068207}. 
Similar composition landscape problems are presented by high-entropy (crystalline) alloys~(HEAs), where the phase stability (at a given temperature) and the challenge of multi-property optimization become evident. 
The high complexity of MGs has in general received a lot of attention from ML approaches \cite{WARD2018102,XIONG2020108378,DOUEST2024411} that help to predict promising compositions with good GFA within a large compositional space. Molecular dynamics~(MD) simulations instead allow 
the computational investigation of MG properties after sample preparation by a given cooling method. The typical application is to compute the mechanical properties of a prepared sample, e.g. in conditions simulating nanoidentation~\cite{fang2018deformation}, deformation protocol dependence~\cite{leishangthem2017yielding}, tensile straining~\cite{bonfanti2018damage}, or shear deformation~\cite{bonfanti2019elementary}.

Moving beyond the progress in predicting GFA, a key question remains: how cooling rates and composition interact to influence the mechanical performance in metallic glasses? 
To answer this question, here, we systematically investigate the link between composition, cooling rate, and mechanical performance within the CuZrAl system, a ternary metallic glass with known good GFA and promising mechanical properties.
This system has recently been explored for the GFA as a function of composition using MD to compute descriptors for that purpose~\cite{li2022data}.
Furthermore, it has been experimentally shown that the addition of small amounts of Al into the CuZr MG increases the mechanical properties, such as strength, ductility, elastic modulus, hardness, and work-hardenability. The optimal amount of Al concentrations seems to vary from 5~\%~\cite{das2005work, yu2008poisson} to exceeding 10~\%~\cite{POLTRONIERI2023119226, cheung2007thermal}.
Additionally, some studies indicate a decrease in the plastic strain is seen with the increase in strength and modulus~\cite{pauly2010transformation}.

In this study we aim to efficiently map the landscape of target mechanical properties using Bayesian optimization and MD simulations. 
Despite the fact that there are fewer compositional degrees of freedom compared e.g. to high entropy alloys~\cite{torsti2024improving, kurunczi2024bayesian}, 
here the sample preparation in simulations is computationally significantly more costly. This is due to the fact that glasses are history-dependent, their properties are significantly influenced by the preparation protocols. 
In order to approach experimental reality in terms of realistic cooling rates, difficult to achieve in numerical simulations, lately Monte Carlo methods have become popular~\cite{parmar2020ultrastable,zhang2022shear,alvarez2023simulated}. In this work we exploit this type of protocol and vary the cooling rate and elemental compositions, to explore its effects on glass properties.

To probe the high-dimensional search space even more efficiently, we employ active learning methods~\cite{ren2023autonomous, khatamsaz2022multi, khatamsaz2023bayesian, mohanty2023machine, siemenn2023fast}. These methods aim to identify optimal compositions with minimal computational cost. 
We use Bayesian optimization~\cite{gelman2013bayesian, eriksson2019scalable, siemenn2023fast, liang2021benchmarking, torsti2024improving, kurunczi2024bayesian}~(BO) with Gaussian Process Regression~\cite{banerjee2013efficient}~(GPR) to explore the search space and to finally exploit the gathered information in finding the optimal composition. Gaussian process regressors require far fewer hyperparameters to be fitted compared for example to neural networks and this translates to significantly smaller datasets needed during BO.
We study \emph{in silico} the CuZrAl metallic glass, and vary its composition around the standard Cu${}_{0.50}$Zr${}_{0.50}$ starting point. A practical reason for this is, in addition to the large interest in this system, the availability of MD potentials.

\begin{figure}[tb!]
\includegraphics[width=\columnwidth]{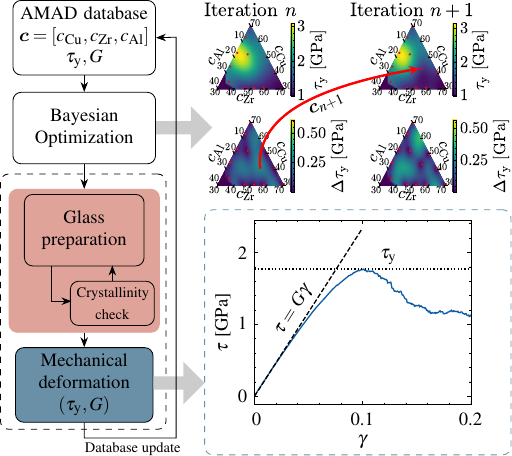}
\caption{
Workflow of the optimization process: data is stored in an online AMAD database from which the BO algorithm determines the next composition~$\bm{c}_{n+1}$, in an MD simulation MGs of this composition are prepared, checked for crystallinity, and sheared. Finally, the database is updated with the yield stress~$\tau_{\rm y}$ and modulus~$G$ determined from the stress-strain curve. The different cooling rates~$\dot{T}$ are run in different loops. The top-right corner illustrates the composition choice made by the BO algorithm: the new composition corresponds to the maximal standard deviation of the yield stress determined by~GPR. The bottom-right corner shows the stress-strain curve and how the two yield point characteristics are derived from it.
}
\label{fig:workflow}
\end{figure}

Specifically, in this work, we utilize an automated protocol~\cite{torsti2024improving} (see Fig.~\ref{fig:workflow}) which combines: an online database, a Bayesian optimization algorithm for the selection of the subsequent composition, and molecular dynamics for the glass preparation and shear simulations to probe the mechanical properties. Our goal is to efficiently map the composition space of CuZrAl glasses in regard to their mechanical properties. In the following, we describe the search process, the effect of the cooling rate, and finally the optimal CuZrAl glass compositions.

\section{Results}
The search process (illustrated in Fig.~\ref{fig:workflow}) starts from a set of four initial compositions---represented by a point in the compositional space $\bm{c} = [c_{\rm Cu}, c_{\rm Zr}, c_{\rm Al}]$ which is just a vector of the atomic composition $c_i$ of the glass---for which the mechanical properties (yield stress~$\tau_{\rm y}$ and the shear modulus~$G$) have been precomputed. These are stored in an online database in the Aalto Materials Digitalisation Platform~(AMAD), which can be accessed both programmatically from the computational cluster (as done in the workflow) or by a human via a web interface.

The main part of the workflow is then the BO algorithm where the mechanical properties at each point in the compositional space are represented by a probabilistic surrogate function, constructed based on known input data (read from the AMAD~database), and is here a Gaussian process regressor. This probabilistic representation is then used for further measurements based on the chosen utility function: one can focus on exploration by examining points where the uncertainty in the probabilistic representation is high, or on exploitation by focusing on points where the value of the mechanical property of interest---here the yield stress~$\tau_{\rm y}$---is expected to be high. Here we have chosen to do a purely exploratory search, and the utility function is simply the standard deviation of the yield stress~$\Delta \tau_{\rm y}$ given by the~GPR. The BO algorithm then gives the next composition to be simulated~$\bm{c}_{n+1}$ (corresponding to the utility function maximum) to the next part of the workflow.

The chosen glass composition is then prepared (by quenching using a hybrid molecular dynamics-Monte Carlo algorithm, see Methods for details) and a shear simulation performed using the LAMMPS~\cite{LAMMPS}. 
The interaction between the atoms is given by Embedded Atom Method~(EAM) 
which has been extensively used for the CuZrAl system~\cite{makinen2024avalanches}.
This is done for 20 random realizations of the composition and the stress-strain curves resulting from the shear simulations are averaged over all the realizations to yield an average stress-strain curve (an example of which is shown in Fig.~\ref{fig:workflow}).

Additionally, the initial glass configuration is checked for crystallinity by computing the pair-correlation function and seeing that there are no sharp peaks indicating crystal or long-range order. The samples shown here were determined to be homogenous and non-crystalline.

The yield stress and shear modulus are then determined from the average stress-strain curve and appended to the~AMAD database. The yield stress corresponds to the maximum stress value in the $\gamma \in (0, 0.2)$ interval, and the shear modulus to the initial slope of the stress-strain curve~(see Fig.~\ref{fig:workflow}).
The workflow will then start another loop.

\begin{figure}[tb!]
\centering
\includegraphics[width=\columnwidth]{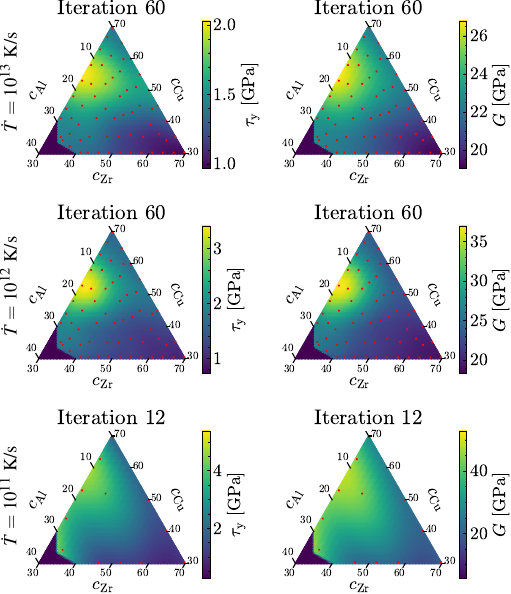}
\caption{The landscape of the mechanical properties obtained at the end of the workflow. The rows indicate different cooling rates~$\dot{T}$ and the columns the mechanical property considered (yield stress~$\tau_{\rm y}$ or shear modulus~$G$.).}
\label{fig:landscape}
\end{figure}

\begin{figure}[tb!]
\centering
\includegraphics[width=\columnwidth]{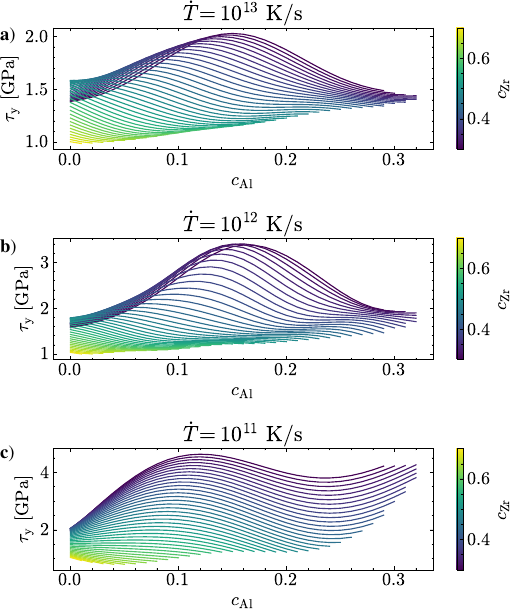}
\caption{The yield stress landscapes of Fig.~\ref{fig:landscape} plotted as a function of the Al~content for various fixed Cu/Zr ratios and for different cooling rates
\textbf{a})~$\dot{T} = 10^{13}$~K/s,
\textbf{b})~$\dot{T} = 10^{12}$~K/s, and
\textbf{c})~$\dot{T} = 10^{11}$~K/s), showing a clear peak at $c_{\rm Al}$ between 10-20~\% at low Zr concentrations.}
\label{fig:projection}
\end{figure}

\subsection{Mapping the mechanical properties in composition space}
As the constraint $\sum_i c_i = 1$ sets up a 2-dimensional manifold in the 3D space, we can simply do the optimization in the 2D space given by two of the compositions, concentrations of Zr and Al.
Additionally, we impose the constraints $0.3 \leq c_{\rm Cu}, c_{\rm Zr} \leq 0.7$ and $c_{\rm Al} \leq c_{\rm Cu}, c_{\rm Zr}$.
After the workflow has run for 60~iterations based on the yield stress $\tau_{\rm y}$ at a cooling rate of $\dot{T} = 10^{12}$~K/s we find the landscape shown in Fig.~\ref{fig:landscape}. Based on the same data we also compute the landscape of the shear modulus~$G$. For comparison, the same sequence of compositions is simulated also for two different cooling rates, a faster~($\dot{T} = 10^{13}$~K/s) and a slower~($\dot{T} = 10^{11}$~K/s) one. All of the resulting landscapes are also shown in Fig.~\ref{fig:landscape} with the exception that the slowest cooling rate is only run for 12~iterations due to the high computational cost.

The yield stress landscapes show a clear maximum at a point corresponding to intermediate Al concentration and low Zr concentration.
The landscapes look pretty similar for different cooling rates, but the scale of the actual yield stress values changes with the cooling rate, the slowly cooled glasses being stronger.
The landscapes of $\tau_{\rm y}$ and $G$ look very similar for all the cooling rates.

The optimal composition for yield stress from GPR is Cu${}_{0.53}$Zr${}_{0.31}$Al${}_{0.16}$, based on the original cooling rate of $\dot{T} = 10^{12}$~K/s. Only minor changes are found for the different cooling rates, 
$\dot{T} = 10^{13}$~K/s giving Cu${}_{0.55}$Zr${}_{0.30}$Al${}_{0.15}$ and 
$\dot{T} = 10^{11}$~K/s giving Cu${}_{0.58}$Zr${}_{0.30}$Al${}_{0.12}$.
The best point found in the dataset (for the original cooling rate) is Cu${}_{0.52}$Zr${}_{0.33}$Al${}_{0.15}$. % Iteration 54

Another way to look at the landscapes is to see how the yield stress varies as a function of the Al concentration (see Fig.~\ref{fig:projection}) for a given concentration of e.g.~Zr.
It is clear that with all the cooling rates the highest yield stresses are achieved at intermediate Al concentrations, at around 10-20~\% Al content in agreement with the experimental picture of the impact of Al addition~\cite{POLTRONIERI2023119226, cheung2007thermal}.
Interestingly, with the slowest cooling case there seems to be a second peak at high aluminium concentrations~(Fig.~\ref{fig:projection}c) which is not present with the faster cooling rates~(Fig.~\ref{fig:projection}a-b). This peak, too corresponds to low Zr concentrations.

\begin{figure}[tb!]
\centering
\includegraphics[width=\columnwidth]{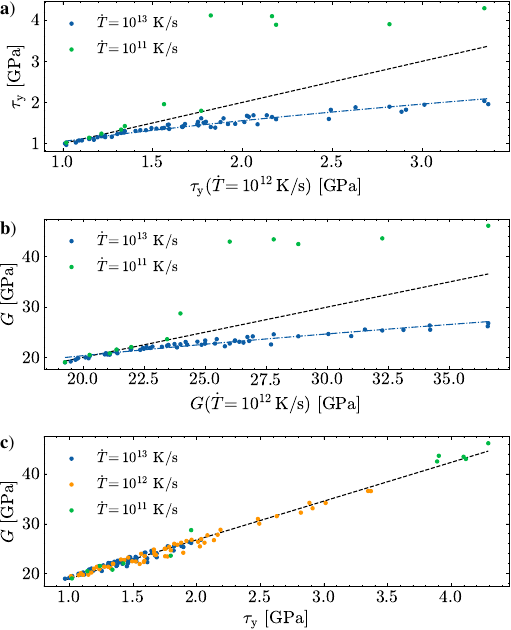}
\caption{
\textbf{a}) The effect of cooling rate on the yield stresses. A dashed line is a square root dependence as a guide to the eye.
\textbf{b}) The effect of cooling rate on the shear modulus. A dashed line is a square root dependence as a guide to the eye.
\textbf{c}) A universal linear relation between the shear modulus and yield stress is found for all cooling rates.
}
\label{fig:scaling}
\end{figure}

The effect of cooling can be seen to change with the mechanical response itself (see Fig.~\ref{fig:scaling}).
The concentrations corresponding to the lowest yield stresses give similar values for all cooling rates~(Fig.~\ref{fig:scaling}a) but when the yield stress increases, the effect of the cooling rate also increases.
The exact same effect is seen in the shear modulus~(Fig.~\ref{fig:scaling}b).

\begin{figure}[tb!]
\centering
\includegraphics[width=\columnwidth]{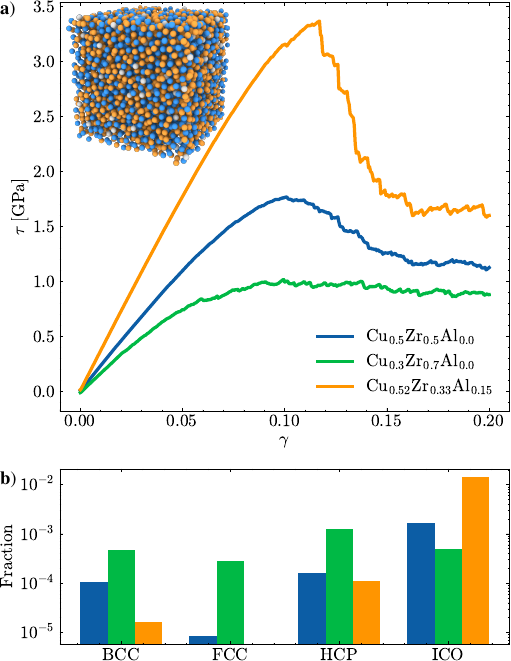}
\caption{\textbf{a})~Example stress-strain curves, illustrating the difference between the starting point Cu${}_{0.50}$Zr${}_{0.50}$, the worst yield stress (Cu${}_{0.30}$Zr${}_{0.70}$) and the best yield stress (Cu${}_{0.52}$Zr${}_{0.33}$Al${}_{0.15}$) at $\dot{T} = 10^{12}$~K/s.
An example configuration of the optimal composition is illustrated in the top-left corner.
\textbf{b})~The fractions of different structures for the compositions shown in panel~\textbf{a}, indicated by the same colors.}
\label{fig:ss}
\end{figure}

The yield stress and shear modulus are strongly related.
A linear relation $G = \tau_{\rm y}/\gamma_0 + G_0$ between them is found (see Fig.~\ref{fig:scaling}c)
with $\gamma_0 = 0.129$ and $G_0=11.24$~GPa. This behavior is seen for all the cooling rates and
arises from the weak dependence of the yield strain~$\gamma_{\rm y}$---defined as the strain corresponding to the yield stress--- on the composition~(see e.g. the stress-strain curves in~Fig.~\ref{fig:ss}).
Generally at the yield point
\begin{equation}
    \gamma_{\rm y} = \frac{\tau_{\rm y}}{G} + \gamma_{\rm p}^{\rm y},
\end{equation} 
where $\gamma_{\rm p}^{\rm y}$ is the plastic strain at the yield point, which can---using the linear relation---be written in the form
\begin{equation}
    \gamma_{\rm p}^{\rm y} = \gamma_{\rm y} - \frac{\tau_{\rm y}}{\tau_{\rm y}/\gamma_0 + G_0}.
\end{equation}
There are two obvious limits: $\tau_{\rm y} \to 0$ where $\gamma_{\rm p}^{\rm y} \to 0$, and
$\tau_{\rm y} \to \infty$ where $\gamma_{\rm p}^{\rm y} \to \gamma_{\rm y} - \gamma_0$.
The value observed for $\gamma_0$ is close to the observed semi-universal $\gamma_{\rm y}$.

\subsection{The optimal glass}
For the original cooling rate of $\dot{T} = 10^{12}$~K/s the optimal composition found in our dataset is Cu${}_{0.52}$Zr${}_{0.33}$Al${}_{0.15}$. Comparing the stress-strain behavior of this composition to the starting point Cu${}_{0.50}$Zr${}_{0.50}$ and the worst composition in the dataset Cu${}_{0.30}$Zr${}_{0.70}$ shows a clear difference in behavior~(see Fig.~\ref{fig:ss}a).
The worst glass shows practically no stress drop, but instead gradually evolves from elastic to plastic behavior. As the yield stress and shear modulus improve, the stress drop becomes more prominent. This shows clearly the impact of composition on mechanical behavior.

Structural explanations for these differences can be studied by examining local structures in these three compositions (see Fig.~\ref{fig:ss}b). There are an inconsequential amount of various types of local crystalline structures, viz BCC, FCC, and HCP for the various compositions~\cite{zemp2015crystal}, which suggests that we are sampling in the ``amorphous-regime" of compositional space. The noticeable change in the mechanical response can be explained by 
the population of the locally favoured icosahedral structures, which are key structural features with long lifetimes and thus provide the "stability" in the supercooled region~\cite{coslovich2007understanding,jakse2008glass}. 
For the various compositions, the population of the icosahedral structures clearly correlates with the dramatic change in the overall mechanical stability. 

\subsection{Search process}
\begin{figure}[tb!]
\centering
\includegraphics[width=\columnwidth]{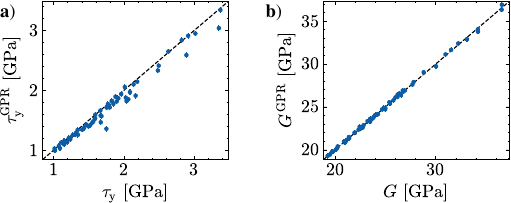}
\caption{
\textbf{a})~The GPR prediction of the yield stress vs. the true value for all the points at iteration 60.
The dashed line indicates the perfect prediction.
\textbf{b})~Same as panel a but for the shear modulus.}
\label{fig:gof}
\end{figure}

The landscapes shown previously~(Figs.~\ref{fig:landscape} and~\ref{fig:projection}) omit the actual measured values and focus on the GPR prediction. Looking at the actual goodness-of-fit at the final iteration~(Fig.~\ref{fig:gof}) shows that the GPR actually fits the data very well.
Interestingly, the fit is slightly worse for the yield stress (Fig.~\ref{fig:gof}a) compared to the shear modulus~$G$~(Fig.~\ref{fig:gof}b). This is likely due to the difficulty of accurately determining the yield stress as the maximum stress achieved at simulation varies strongly, which affects the maximum stress in the average stress-strain curves. The shear modulus is much more robust against this type of stochastic behavior.

Good glasses are found really early in the search process, the second best yield stress (3.01~GPa) being achieved already at iteration 6.
The best yield stress~(3.36~GPa) corresponds to iteration~54, and the third (2.94~GPa) and fourth (2.92~GPa) to iterations~23 and~29, respectively. This shows that if we would choose to do exploitative optimization instead of the exploratory search used here, the number of iterations needed would likely be much smaller than the~60 used here.

\begin{figure}[tb!]
\centering
\includegraphics[width=\columnwidth]{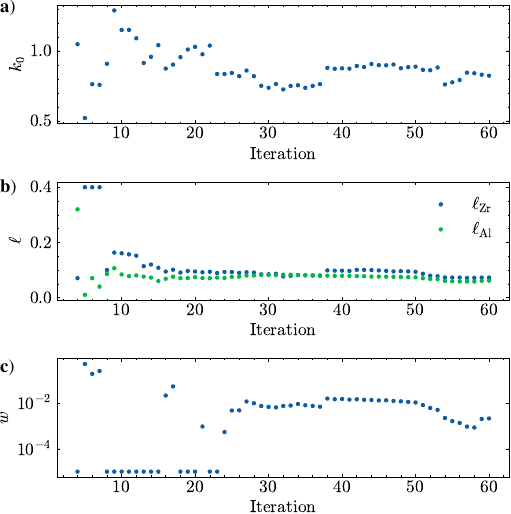}
\caption{The evolution of the hyperparameters
\textbf{a}) kernel strength~$k_0$, 
\textbf{b}) kernel lengthscales~$\ell_i$, and
\textbf{c}) white noise kernel strength~$w$
as a function of the iteration number.}
\label{fig:hyperparams}
\end{figure}

At each iteration of the search, the GPR kernel hyperparameters ($k_0$, $\ell_i$, and $w$, see Methods for details) are optimized. 
The evolution of these with the iteration number~(Fig.~\ref{fig:hyperparams}) shows that after around ten iterations the hyperparameter values seem to converge to roughly correct numbers.
This is consistent with behavior seen in Figs.~\ref{fig:landscape} and~\ref{fig:projection} where the slowest cooling rate ($\dot{T} = 10^{11}$~K/s) shows similar behavior as the other cooling rates already at iteration~12 (compared to the iteration~60 for the others).
The only exception is the white noise level~$w$ which stays at the minimum value~($w=10^{-5}$). In the next ten iterations, the noise level~$w$ rises, and the other hyperparameters converge to $k_0 \approx 0.8$ and $\ell_{i} \approx 0.1$. The length scale is practically the same for both Al and Zr.
The hyperparameter values stay there for the rest of the iterations with only minor fluctuations around the values achieved at iteration~30.
The white noise level~$w$ stays very low at all the iterations, signifying very low errors in the GPR fit.

\section{Discussion}
We have implemented an automated loop for exploring the compositional space of CuZr(Al) metallic glasses, studying also various cooling rates. This is done by utilizing Bayesian optimization, cooling in silico and molecular dynamics shear simulations.

The landscape of the system (with some constraints as to the Cu/Zr content) turns out to be dominated by one maximum "peak" with the two implications of non-equiatomic Cu/Zr ratio and an optimal Al content. This appears to be in line with the somewhat thin experimental understanding. Likewise, the idea that extra Cu content or a non-equimolar Cu/Zr composition is good for the GFA has been made with a similar qualitative twist of Cu dominance \cite{li2022data}. The effect of the cooling rate turns out to be very relevant because the optimal compositions can indeed be controlled even more by slower cooling.

Presently, we do not seek the fundamental relationship between the structure and the mechanical properties and the structural signatures, controlling the elastic modulus or yield stress. However, the question alluded to above of GFA optimization is clearly related to these. The same also applies to the eventual presence of short- or medium-range order that could be varied with cooling.

In summary, Bayesian approaches seem to be promising for the exploration of metallic glasses due to the "small data" nature of these workflows (we ran 60~iterations compared to the 813~possible grid points). We would in particular stress the applications to experiments, where one thus could hope to avoid complex high-throughput setups by a design of the experimental search/mapping process. For our \emph{in silico} case we have selected a serial BO approach, but in experimental adaptations, it would also be an avenue to look at batch variants \cite{pmlr-v51-gonzalez16a} of Bayesian search so as to speed up the research compared to the serial exploration. The same also applies to multi-target optimization \cite{Alvia2024multiobj} and property mappings such as GFA and mechanical properties.

\section{Methods}
The Bayesian Optimization~(BO) is done here using Gaussian Process Regression~\cite{banerjee2013efficient}~(GPR), implemented in \textsc{scikit-learn} software~\cite{pedregosa2011scikit}, where the desired quantity $y$ for each point $\bm{c}$ in the search space is represented by a Gaussian distribution with mean $\mu_y(\bm{c})$ and standard deviation~$\Delta y(\bm{c})$.
The points in the search space are given by vectors $\bm{c} = [c_{\rm Zr}, c_{\rm Al}]$ reduced into 2D, as the $\sum_i c_i = 1$ constraint makes this possible.
The kernel used in the GPR is based on the standard anisotropic radial basis function~$k_{\rm RBF}$, characterized by a kernel strength~$k_0$ and two length scales $\ell_i$ corresponding to the $i$th component of the compositions. 
To account for the variation in the yield stress due to the specific atomic configuration a white noise kernel, $k_{\rm WN}\left[ y(\bm{c}), y(\bm{c}') \right] = w \delta\left(|\bm{c}-\bm{c}'|\right)$ where $\delta$ is the Dirac delta function and the norm the Euclidean distance, is added. The total kernel is then given by the covariance $\mathrm{cov}\left[y(\bm{c}), y(\bm{c}') \right] = k_{\rm RBF} + k_{\rm WN}$.
The kernel hyperparameters $w$, $k_0$ and $\ell_i$ optimized in the GPR algorithm are chosen to have the following initial values: $w$ is initially the standard deviation of the yield stress in the current dataset, $k_0$~four times this value, and the length scales $\ell_i$ are bounded by the minimum and maximum spacing of the datapoints in the dataset~\cite{miranda2022predicting, torsti2024improving} (i.e. the minimum and maximum difference of $c_i$ in the dataset), and initialized to the mean of these two values. 
The white noise level~$w$ is restricted by $w \geq 10^{-5}$ to avoid numerical issues.

The output used in the active learning part of the workflow is the yield stress $\tau_{\rm y}$. In addition to this, we use the shear modulus~$G$ data to do similar GPR fitting, although this is not involved in the active learning part.
Finally, the same sequence of compositions is simulated also for two different cooling rates, again without the active learning component.
The optimization starts from four initial inputs 
Cu${}_{0.50}$Zr${}_{0.50}$, Cu${}_{0.70}$Zr${}_{0.30}$, Cu${}_{0.30}$Zr${}_{0.70}$, and Cu${}_{0.34}$Zr${}_{0.34}$Al${}_{0.32}$, representing three commonly studied CuZr glasses at the edges of the search space, and a sample with the maximum Al-content.
After this, the next point is picked based on the highest value of the utility function.
We have here chosen to perform a purely exploratory search, and use as the utility function the standard deviation of the yield stress~$\Delta \tau_{\rm y}$.

Optimization in a grid can run into a situation where the maximum of the utility function is achieved at a point already previously visited.
However, this is much less likely in the purely exploratory search. We have chosen as the next point the point corresponding to the maximum utility function, in the set of points not previously visited.
We run the search loop for 60~iterations, which gives a fairly good coverage of a 2D~surface. For the slowest cooling rate, we have stopped the run after 12~iterations, due to the high computational cost.\\

To prepare the glasses and test the mechanical properties of each composition we perform MD simulations using LAMMPS~\cite{LAMMPS}.
The system size used here is 6000 atoms and a cubic box with periodic boundary conditions is used.
Glass preparation is done using a hybrid Molecular Dynamics-Monte Carlo~(MD+MC) scheme under the variance-constrained semi-grand canonical ensemble~(VC-SGC)~\cite{sadigh2012scalable}. This VC-SGC MC scheme allows exploring the configurational degrees of freedom by randomly selecting an atom and attempting to change its type, while also calculating the corresponding energy and \textit{concentration} changes. It allows to targeting of specific concentration ranges while maintaining a fixed total number of particles and the volume. Acceptance of these transmutations follows the Metropolis criterion, ensuring the preservation of detailed balance. 
On the other hand, the relaxation processes are accounted for by the MD integration steps. To maintain the desired concentration within the system \cite{sadigh2012scalable}, we set the variance parameter~$\kappa$=10$^3$. The differences in chemical potential relative to Zr using hybrid MD+MC~simulations under the semi-grand canonical ensemble at a temperature of 2000~K and the specific set of parameters that minimize the composition errors in relation to the desired concentration can be found in Ref.~\cite{alvarez2023simulated}.
The hybrid scheme is also used for ZrCu by Ref.~\cite{zhang2022shear}.
The interaction between the atoms is given by the embedded atom method~(EAM) interatomic potential and parameters~\cite{cheng2009atomic}.
The quenching is performed starting from molten metals at a high temperature well above the melting point using a fixed cooling rate~$\dot{T}$ at fixed pressure. In the MD+MC~scheme a MC~cycle consisting of $N$ attempts is performed every 20~MD~steps. The glasses are cooled from 2000~K to 300~K,
using one of the three cooling rates ($10^{11}$, $10^{12}$, or $10^{13}$~K/s) as explained in the main text.

The structural analysis was done using \textsc{ovito} software~\cite{stukowski2009visualization}. The fractions of atoms participating in face-centered-cubic~(FCC), hexagonal close-packed~(HCP), body-centered-cubic~(BCC) and icosahedral~(ICO) structures were determined while considering the nearest geometric neighbours. The crystallinity check involved observing the fractions of the crystalline structures, which are clearly not significant.

The shear simulations are performed in three dimensions with periodic boundary conditions, using the glass configurations cooled to~$T=300$~K. The simulation box is incrementally sheared along the $x$-direction with respect to the $y$-direction by~$\delta\gamma = 10^{-4}$, and at each strain increment, MD simulations are performed for a duration of 1~ps. Throughout the shearing process, we record the stress-strain response, which provides a detailed stress-strain curve for analysis.
The mechanical property we focus on is the yield stress~$\tau_{\rm y}$, which is determined from the stress-strain curve averaged over 20~realizations of the atomic configurations as the maximum of the shear stress. Also, the shear modulus~$G$, determined by averaging the positive slopes of the stress-strain curve for the first 1~\% of strain in each realization of the disorder, is recorded. See Fig.~\ref{fig:workflow} for an illustration of the stress-strain curve analysis.\\

The full workflow (illustrated in Fig.~\ref{fig:workflow}) starts with a database stored in the Aalto Materials Digitalization Platform~(AMAD), which has an initial set of inputs and outputs. 
The database is read automatically by the BO~algorithm, which performs the~GPR in a grid with a concentration step $\Delta c = 0.01$ and boundaries $0.30 \leq c \leq 0.70$, which is the window in which we let the concentrations to vary. We also impose the additional conditions $c_{\rm Al} < c_{\rm Cu}$ and $c_{\rm Al} < c_{\rm Zr}$.
The algorithm outputs a point that corresponds to the highest value of the utility function~($\Delta \tau_{\rm y}$) in the points that have not been visited before. 
The composition corresponding to this point is given as an input to the shear simulation and after the yield stress and shear modulus are determined, the composition and the corresponding yield stress are written back to the AMAD database. This process is automatically iterated.
\section*{Data Availability}
The datasets used and/or analysed during the current study available from the corresponding author on reasonable request
\section*{Code Availability}
The underlying code for this study is not publicly available but may be made available to qualified researchers on reasonable request from the corresponding author.
\begin{acknowledgments}
A.D.S.P, S.B. and M.J.A. are supported by the European Union Horizon 2020 research and innovation program under grant agreement no.~857470 and from the European Regional Development Fund via the Foundation for Polish Science International Research Agenda PLUS program grant No.~MAB PLUS/2018/8.
M.J.A. acknowledges support from the
Academy of Finland (361245 and 317464) and from the Finnish Cultural Foundation. 
S.B. acknowledges support from the National Science Center in Poland through the SONATA BIS grant DEC-2023/50/E/ST3/00569.
T.M., and M.J.A. acknowledge support from the FinnCERES flagship (151830423), Business Finland (211835, 211909, and 211989), and Future Makers programs.
The authors acknowledge the computational resources provided by the Aalto University School of Science “Science-IT” project.
\end{acknowledgments}
\section*{Author Contributions}
T.M., A.D.S.P,
S.B., and M.J.A. conceptualized the research and developed the methodology.
T.M. carried out the simulations.
T.M. and A.D.S.P. performed the data analysis.
All authors contributed to the writing of the manuscript.
\section*{Competing Interests}
The authors have no competing interests to declare.\\
%\bibliography{biblio}% Produces the bibliography via BibTeX.

\end{document}